\PassOptionsToPackage{authoryear,round}{natbib}
\PassOptionsToPackage{table}{xcolor}

\documentclass[11pt,a4paper]{article}

\usepackage[a4paper,margin=1in,includefoot]{geometry}
\usepackage[T1]{fontenc}
\usepackage[utf8]{inputenc}
\usepackage{lmodern}

\usepackage{amsmath,amssymb}

\usepackage{graphicx}
\usepackage{xcolor}
\usepackage{float}
\usepackage{pdfpages}
\usepackage{adjustbox}
\usepackage{comment}

\usepackage{algorithm}
\usepackage{algorithmic}

\usepackage{multirow}
\usepackage{booktabs}
\usepackage{threeparttable}
\usepackage{enumitem}
\usepackage{tabularx,array}
\usepackage{tikz}

\newcolumntype{L}[1]{>{\raggedright\arraybackslash}p{#1}}
\newcolumntype{C}{>{\centering\arraybackslash}X}

\renewcommand{\arraystretch}{1.06}

\usepackage{subcaption}
\usepackage{natbib}
\bibpunct{(}{)}{;}{a}{}{,}
\usepackage[hidelinks]{hyperref}
\usepackage{url}

\newcommand{\orcidlink}[1]{} 

\newcommand{\AJSAbstractText}{}
\newcommand{\AJSKeywordsText}{}

\newcommand{\Abstract}[1]{\gdef\AJSAbstractText{#1}}
\newcommand{\Keywords}[1]{\gdef\AJSKeywordsText{#1}}

\newcommand{\Plainauthor}[1]{}
\newcommand{\Plaintitle}[1]{}
\newcommand{\Shorttitle}[1]{}
\newcommand{\Pages}[1]{}

\definecolor{indianred1}{HTML}{FF6A6A}
\definecolor{indianred3}{HTML}{CD5555}
\definecolor{lightpink2}{HTML}{EEA2AD}

\definecolor{coral1}{HTML}{FF7256}
\definecolor{coral3}{HTML}{CD5B45}
\definecolor{darkorange}{HTML}{FF8C00}

\definecolor{gray70}{HTML}{B3B3B3}
\definecolor{gray40}{HTML}{666666}
\definecolor{gray24}{HTML}{3D3D3D}

\definecolor{nfortyA}{HTML}{C89414}
\definecolor{nfortyB}{HTML}{CCBC72}
\definecolor{nfortyC}{HTML}{8D8A0F}

\definecolor{nfiveA}{HTML}{C27BB0}
\definecolor{nfiveB}{HTML}{8B5F92}
\definecolor{nfiveC}{HTML}{B695C0}

\definecolor{steelblue4}{HTML}{36648B}
\definecolor{steelblue3}{HTML}{4F94CD}
\definecolor{lightblue}{HTML}{ADD8E6}

\newcommand{\chip}[1]{%
  \begingroup\setlength{\fboxsep}{0pt}%
  \colorbox{#1}{\rule{0pt}{0.9em}\rule{0.9em}{0pt}}%
  \endgroup
}

\newcommand{\Nchips}[1]{%
  \begingroup\count0=#1\relax
  \ifnum\count0=0
    \chip{white}\chip{white}\chip{white}\chip{white}\chip{white}\chip{white}%
  \else\ifnum\count0=20
    \chip{indianred3}\chip{indianred1}\chip{lightpink2}\chip{gray70}\chip{gray40}\chip{gray24}%
  \else\ifnum\count0=40
    \chip{nfortyA}\chip{nfortyB}\chip{nfortyC}\chip{gray70}\chip{gray40}\chip{gray24}%
  \else\ifnum\count0=80
    \chip{steelblue4}\chip{steelblue3}\chip{lightblue}\chip{gray70}\chip{gray40}\chip{gray24}%
  \else\ifnum\count0=200
    \chip{coral1}\chip{coral3}\chip{darkorange}\chip{gray70}\chip{gray40}\chip{gray24}%
  \else\ifnum\count0=500
    \chip{nfiveA}\chip{nfiveB}\chip{nfiveC}\chip{gray70}\chip{gray40}\chip{gray24}%
  \fi\fi\fi\fi\fi\fi
  \endgroup
}

\title{Joint return levels of maximum temperature and minimum relative humidity by combining copulas with an extreme value framework for bimodal data}

\author{%
\begin{tabular}{c}
Beatriz G da Cruz Albernaz$^{1,*}$ \quad
Cira E G Otiniano$^{1,*}$ \quad
Carolyne Soares de Brito$^{1}$
\\[0.2em]
Fidel E C Morales$^{2}$ \quad
Enzo Porto Brasil$^{1}$
\end{tabular}
\\[1cm]
{\small
$^{1}$Department of Statistics, University of Brasília, Brazil
\\
$^{2}$Department of Statistics, Federal University of Rio Grande do Norte, Brazil
}
\\[1em]
\scriptsize{$^{*}$Corresponding authors: Cira Otiniano (cira@unb.br); Beatriz Albernaz (beatrizgomesalbernaz@gmail.com)
}
}

\Abstract{%
Climate change has become a growing concern, particularly in regions experiencing increasingly frequent extreme events. In Brasília, the capital of Brazil, significant shifts in climate patterns have drawn attention, including episodes of intense heat, unusually cold weather, and prolonged dry periods, often accompanied by wildfires. These phenomena directly affect the population and local ecosystems, requiring detailed analyses to understand their causes and effects.
In this work, the return period for the joint distribution of maximum air temperature and minimum relative humidity in Brasília is determined using a copula-based approach. First, the dependence structure between the variables is modeled through a copula, while the marginal distributions are fitted using the novel methodology for modelling extreme values in complex systems.
Different copula families, including rotated versions, were evaluated using information criteria, leading to the selection of a model that adequately captured the dependence structure, including asymmetry and tail behavior. The selected model also reproduced the multimodal pattern observed in the joint density, a feature consistent with the possible presence of multiple climate regimes.
In the context of climate data modelling, the bivariate return level was determined using a conditional approach to assess the occurrence of extreme scenarios characterized by high temperatures combined with low relative humidity. This analysis makes it possible to quantify the expected frequency of such events, providing valuable information for environmental risk monitoring, the planning of preventive measures, and the development of climate change adaptation strategies in the region.
}

\Keywords{EVT; BGEV; Copula; Return levels; Bimodal data.}

\usepackage{fancyhdr}

\newcommand{\PreprintLine}{\textit{Preprint. \today}}

\fancypagestyle{firstpage}{%
  \fancyhf{}
  
  \fancyfoot[L]{\small \PreprintLine}
  \fancyfoot[C]{\small \thepage} 
  \fancyfoot[R]{}
}

\makeatletter
\renewcommand{\maketitle}{%
  \thispagestyle{firstpage}%
  \vspace*{-1.2em}%
  \hrule height 0.6pt%
  \vspace{0.9em}%
  \begin{center}
    {\Large\bfseries \@title\par}
    \vspace{1.0em}
    {\normalsize \@author\par}
  \end{center}

  \@thanks

  \vspace{0.9em}%
  \hrule height 0.6pt%
  \vspace{1.2em}%
}
\makeatother

\date{} 

\begin{document}

\maketitle

\begin{abstract}
\AJSAbstractText
\end{abstract}

\vspace{0.8em}
\noindent\textit{\textbf{Keywords }} \AJSKeywordsText

\vspace{2em}


\section{Introduction}
Climate change observed in recent decades has been accompanied by an increase in the frequency and intensity of extreme events across many regions, making their analysis a central concern in environmental and statistical research. Findings reported in the literature show that this process can be observed across diverse geographical settings (\citeauthor{IPCC2013} \citeyear{IPCC2013}; \citeauthor{WMO2026}, \citeyear{WMO2026}). In Ethiopia, significant changes in temperature and precipitation patterns have been recorded, accompanied by increasingly recurrent droughts \citep{ETIOPIA}; India has experienced more frequent, intense, and prolonged heatwaves \citep{INDIA}; the Antarctic Peninsula has exhibited marked warming trends and climate variability \citep{ANTARTICA}; and regions across sub-Saharan Africa have shown high vulnerability to climatic changes, with direct impacts on the socioeconomic dynamics of the region \citep{AFRICA}.

In Brazil, these changes have been particularly evident. In 2024, for example, the country experienced several extreme events, including floods in the South region, wildfires in the Pantanal area and the Cerrado savannas, severe droughts in the Amazon, and heatwaves in the Central-West, Southeast, and South regions \citep{marengo2025clima}. Such phenomena directly affect human health, ecosystems, and socioeconomic dynamics, underscoring the importance of statistically analyzing extreme climate events within specific regional contexts \citep{Clima_Brasil}.
Among climate-related studies, the work of \citet{GEV_BRASIL} is particularly noteworthy. The authors analyzed annual maximum temperatures in 27 Brazilian state capitals from 1960 to 2020, finding significant increases in 24 of them and identifying abrupt changes in the behavior of the time series. In addition to the empirical analysis, they modeled the data using the Gumbel (GUM) and Generalized Extreme Value (GEV) distributions, demonstrating the usefulness of these approaches for modelling heatwaves, droughts, and wildfires.

The First National Assessment Report of the Brazilian Panel on Climate Change \citep{PBMC2014} emphasizes that the impacts of climate change vary across Brazilian regions and biomes. In the Cerrado, including the Federal District, rising temperatures, decreasing precipitation, and longer dry seasons are projected, all of which increase vulnerability to wildfires and may intensify carbon emissions into the atmosphere.

In this context, the Federal District has specific climatic characteristics that make it particularly relevant to studies of climate extremes. Traditionally known for its dry climate and moderate temperatures, Brasília has experienced increasingly intense climate extremes in recent years, including record-high temperatures, extended periods of low relative humidity, prolonged droughts, and episodes of heavy rainfall \citep{Menezes2016}. These events have directly affected the daily life of the population and the environmental quality of the region, reinforcing the need for quantitative studies to improve our understanding of the magnitude and recurrence of such phenomena.
The recurrence of these conditions in the region is also reflected in the frequent warnings issued by the Brazilian National Institute of Meteorology (INMET). A report published by Correio Braziliense highlighted the possibility of relative humidity falling to approximately 10\% in the Federal District \citep{Correio2025}.

Different statistical approaches have been applied to the study of climate variables in the Federal District. \citet{Bayer2015} modeled and forecast relative humidity in Brasília using the beta autoregressive moving average ($\beta$ARMA) model, highlighting the importance of monitoring and forecasting this variable in fields such as public health, water resources, and climate studies. Another methodology for modelling extreme data was employed by \citet{Cordeiro2020}, who used bivariate models to assess the dependence among extreme events involving maximum temperature, low relative humidity, maximum wind speed, and maximum solar radiation in the Federal District, demonstrating the joint occurrence of these variables under severe weather conditions.


However, the minimum relative humidity extremes analyzed in the present study exhibited heterogeneous and bimodal behavior. Conventional marginal fits were therefore unsatisfactory because the GEV distribution is unimodal. Similarly, traditional bivariate models with unimodal margins were unable to adequately represent the observed joint structure.

Despite the relevance of these climate variables to the Federal District, studies devoted to the probabilistic analysis of the dependence between temperature and relative humidity in the region remain scarce. To the best of our knowledge, no previous study has jointly modeled these variables using a bivariate distribution with a bimodal margin.


Accordingly, this study aims to develop an approach capable of adequately modelling extreme climate data exhibiting bimodal behavior in both univariate and bivariate settings. More specifically, the proposed approach involves constructing a bivariate model that combines several copula families with potentially bimodal extreme-value marginal distributions, such as those proposed by \citet{otiniano2025revised}.

As a second contribution of this study, the new bivariate bimodal approach is used to determine the return level $(x,y)$ of maximum temperature $X$ and minimum relative humidity $Y$ for specified return periods $T$, considering the joint probability $P(X>x, Y<y)$. This approach is motivated by the need to characterize the simultaneous occurrence of extreme maximum temperature $X$ and minimum relative humidity $Y$.

The remainder of this paper is organized as follows. Section 2 presents the proposed methodology for modelling bivariate extreme data. Section 3 begins with an exploratory analysis of different datasets comprising climate variables from the Federal District. Pairs of these variables are then fitted using the proposed model, and their bivariate return levels are calculated for different return periods. Finally, we present our conclusions and discuss possible directions for future research.

\section{Methodology}\label{metodologia}
In this section, we propose a new approach to modelling bivariate extreme data.

\subsection{Bivariate Extreme-Value Model with Bimodal Margins}
Our proposed approach is based on Sklar’s theorem \citep{sklar1959fonctions}. Thus, we assume that the joint distribution of a random vector $(X,Y)$ is given by a joint distribution function $\mathbf{F}$, obtained by combining a copula $C$ with the marginal distributions $F_1$ and $F_2$,

\begin{equation}\label{F}
\mathbf{F}(x_1, x_2) = C(F_1(x_1), F_2(x_2)).
\end{equation}

In this study, we consider the Independence, Gaussian, Student-t, and Frank copula families, as well as the Archimedean Clayton, Gumbel, and Joe families, including their respective rotated versions at $90^\circ$, $180^\circ$, and $270^\circ$. In addition, the BB1, BB6, BB7, BB8, and Tawn families (Types I and II) are also evaluated, together with their rotated versions when available. The most appropriate copula is selected based on the Akaike Information Criterion (AIC) and the Bayesian Information Criterion (BIC). The selected model is the one with the highest log-likelihood and the lowest AIC and BIC values.

In the context of extreme-value modelling, it is common to assume that the marginal distributions of $X$ and $Y$ follow the Generalized Extreme Value (GEV) distribution. In this study, however, $F_1$ and $F_2$ are assumed to follow the Bimodal Generalized Extreme Value (BGEV) distributions proposed by \citet{otiniano2025revised}. These distributions constitute a bimodal extension of the GEV distribution, whose cumulative distribution function is given by

\begin{eqnarray}\label{Fmarg}
F(t;{\xi,\mu,\sigma,\delta}) =
\begin{cases}
\exp\left[-\left[1 + \xi \left(\frac{(t - \mu) |t - \mu|^{\delta}}{\sigma}\right)\right]^{-1/\xi}\right], & \xi \neq 0, \\
\exp\left[-\exp\left[-\frac{(t - \mu) |t - \mu|^{\delta}}{\sigma}\right]\right], & \xi = 0,
\end{cases}
\end{eqnarray}
 where $\sigma>0$, $\delta \geq 0$, and $\xi \in \mathbb{R}$ are shape parameters, and $\mu \in \mathbb{R}$ is a location parameter.

 In addition to being a shape parameter, $\delta$ determines whether the distribution is unimodal, when $\delta=0$, or bimodal, when $\delta>0$, whereas $\xi$ determines whether the distribution has light or heavy tails \citep{otiniano2025revised}. 

The inverse function of $F$ is defined as:

\begin{equation}\label{Finv}
    F^{-1}(t) =
\begin{cases}
\mu + \operatorname{sign}\left(\frac{\sigma}{\xi}[(-\log(t))^{-\xi} - 1]\right) \left|\left(\frac{\sigma}{\xi}[(-\log(t))^{-\xi} - 1]\right)\right|^{\frac{1}{\delta+1}}, & \xi \neq 0, \\
\mu + \operatorname{sign}(-\sigma \log(-\log(t))) |-\sigma \log(-\log(t))|^{\frac{1}{\delta+1}}, & \xi = 0,
\end{cases}
\end{equation}

\noindent  where
 $$\operatorname{sign}(t) =
\begin{cases} 
1, &  t > 0, \\
0, &  t = 0, \\
-1, &  t < 0.
\end{cases}$$

From Equation (\ref{F}), if $F_1(X)=U$ and $F_2(Y)=V$, then

\begin{align}
\label{Cuv}
C(u,v)
  &= \mathbf{F}\!\left(F_1^{-1}(u),F_2^{-1}(v)\right) \notag\\
  &= \mathbf{F}(x,y),
\end{align}

\noindent where the inverse marginal functions are given by Equation~(\ref{Finv}).
The formulation in (\ref{Cuv}) is widely used in the analysis of bivariate extreme events because it allows marginal distributions suitable for extremes to be combined with different dependence structures \citep{salvadori2007extremes}. An alternative for marginal distributions with more than one mode is to use mixtures of extreme-value distributions; however, this approach makes the estimation of the joint model parameters computationally demanding.

The copula $C$ may be selected from several well-established copula families, including elliptical, Archimedean, extreme-value, and BB. Throughout this paper, we adopt the notation for copula families described in \cite{joe1996estimation}. 

A key feature of the proposed distribution (\ref{F}) with bimodal margins $F_1$ and $F_2$ is its ability to exhibit more than one mode in the joint density. This property makes the proposed model substantially more flexible than standard copula-based constructions with unimodal margins.

\subsection{Return levels}
According to \citet{shiau2003return}, suppose that an extreme event occurs whenever a random variable $X$, with distribution function $F_1$, exceeds a specified extreme quantile $x_T$, known as the return level of $X$. The time between consecutive occurrences of the event $[X>x_T]$ is denoted by $T_X$, and the return period associated with the event $[X>x_T]$ is defined as the expected value of $T_X$, $E(T_X)=T_1$, given by:
\[
T_1=\frac{E(L)}{1-F_1(x_T)},
\]
where $L$ denotes the time interval between the occurrence of any two successive events of the type $[X>x_T]$. From this expression, the probability of exceeding the extreme return level $x_T$ is given by
\[
P(X > x_T) = \frac{E(L)}{T_1}.
\]
For the annual series, the mean interarrival time is typically assumed to satisfy the condition $E(L)=1$. In the present study, however, the series is treated as a partial series because the block maxima (minima) approach is used to obtain observations that can be regarded as independent and identically distributed. Further details on the specification of $E(L)$ are provided in the application section.

The return period of a variable is useful when only one extreme random event is relevant to the design criterion under consideration. However, when two variables $(X,Y)$ are involved, different definitions of joint extreme events may be considered, depending on the nature of the variables under study \citep{salvadori2007extremes}. In many studies, an extreme event is defined as the simultaneous occurrence of high values of both variables, with the return level obtained from the joint exceedance probability $P(X>x, Y>y)$. Alternatively, some studies consider the joint occurrence of low values, defining the extreme event through the probability $P(X<x, Y<y)$. The construction of these return levels can be found in studies such as \cite{graler2013multivariate} and \cite{vandenberghe2012joint}. These different definitions correspond to distinct regions in the bivariate space and should be selected according to the physical phenomenon under investigation \citep{salvadori2007extremes,joe2014dependence}.

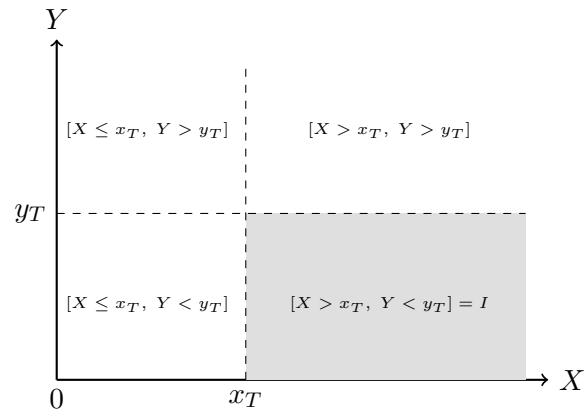
\begin{figure}[htbp]
\centering
\begin{tikzpicture}[scale=1]

\draw[->, thick] (0,0) -- (6.5,0) node[right] {$X$};
\draw[->, thick] (0,0) -- (0,4.5) node[above] {$Y$};

\fill[gray!25] (2.5,0) rectangle (6.2,2.2);

\draw[dashed] (2.5,0) -- (2.5,4.2);
\draw[dashed] (0,2.2) -- (6.2,2.2);

\node[below] at (0,0) {$0$};
\node[below] at (2.5,0) {$x_T$};
\node[left] at (0,2.2) {$y_T$};

\node at (1.2,3.3) {\tiny{$[X\le x_T,\;Y>y_T]$}};
\node at (4.4,3.3) {\tiny{$[X>x_T,\;Y>y_T]$} };

\node at (1.2,1.0) {\tiny{$[X\le x_T,\;Y<y_T]$ }};
\node at (4.4,1.0) {\tiny{$[X>x_T,\;Y<y_T]=I$ }};

\end{tikzpicture}
\caption{Region of interest highlighted in gray.}
\label{fig:quadrantes}
\end{figure}

In the present study, the event of interest is defined as
\[
I=[X>x,Y<y],
\]
which represents the simultaneous occurrence of high values of the variable $X$ and low values of the variable $Y$. Figure \ref{fig:quadrantes} illustrates the region corresponding to this joint event.

In region $I$, extreme values of two random variables occur simultaneously, such as maximum temperature values ($X$) and minimum relative humidity values ($Y$).

Let $C$ be the copula associated with the joint distribution of $(X,Y)$, defined in (\ref{F}). Then, the probability of event $I$, $p=P(X>x_T, Y<y_T)$, is given by

\begin{eqnarray*}
    P(X>x_T, Y<y_T) &=& F_Y(y_T) - C(F_X(x_T),F_Y(y_T))
    \nonumber\\
               &=&v- C(u,v).
\end{eqnarray*}
Thus, the return period associated with the extreme event $I$ is denoted by $T_2$ and satisfies

\begin{eqnarray*}
    p&=&\frac{E(L)}{T_2},
\end{eqnarray*}
where $L$ denotes the time interval between the occurrence of any two successive events of type $I$. For a given return period $T_2$, the return levels $(x_T , y_T)$ are solutions to the equation

\begin{eqnarray}\label{nivelreturn}
[v- C(u,v)]|_{u=F_X(x_T), v=F_Y(y_T)}=\frac{E(L)}{T_2}.  
\end{eqnarray}

Note that, for a fixed value of $T_2$, the pair $(x_T,y_T)$ satisfying (\ref{nivelreturn}) is not unique. Instead, the solutions form a return-level curve. To determine this curve numerically, a grid of values $u_1,u_2,\ldots,u_n$ is defined over the interval $(0,1)$. Then, for each value $u_i$, the corresponding value $v_i$ satisfying (\ref{nivelreturn}) is determined. Thus, the values of $(x_i, y_i)$ defining the return-level curve are obtained from $x_i=F_X^{-1}(u_i)$ and $y_i=F_Y^{-1}(v_i)$.

Although the bivariate return levels associated with a given return period form a curve, in some applications it is useful to select a representative point on this curve for interpretation or comparison across different return periods.

In this study, among all the solutions obtained for a given return period, the point $x_T$ corresponding to the largest value of $u$ is selected. That is, we select the temperature value $X$ associated with the largest cumulative probability, $u = F_X(x)$. The value $y_T$ satisfying the condition $P(X>x, Y<y)=p$ is then determined. This strategy emphasizes events characterized by more extreme values of the variable $X$ while preserving the dependence structure modeled by the copula \citep{salvadori2007extremes,joe2014dependence}.


\subsubsection{Procedure for Calculating the Return Level}
\label{subsec:procedimento}

The return levels associated with event $I$ are determined through the following steps:

\begin{enumerate}
\item Selecting extremes (maxima of $X$ and minima of $Y$) from blocks of size $N$ such that the resulting extreme-value series exhibit low correlation (see \cite{otiniano2025revised}).
\item Fitting each extreme-value series ${x_i}$ and ${y_i}$ using the distribution in (\ref{Fmarg}); $X\sim F_X$ and $Y\sim F_Y$;

\item Transforming the extreme-value data to the unit interval using $F_X(x_i)=u_i$ and $F_Y(y_i)=v_i$;

\item  Selecting the parametric copula that best fits the data $\{(u_i,v_i)\}$ based on the lowest AIC and BIC values;

\item Defining the probabilities associated with the return period as $p=E(L)/T$, with $E(L)=N/365$ when $T$ is measured in years, and $E(L)=N$ when $T$ is measured in days, where $N$ is the block size;
\item Generating a grid of values for $u$;
\item Determining the value of $v$ that solves Equation (\ref{nivelreturn}), $C(u,v)=v-p$;

\item Transforming the results back to the original scale using the inverse marginal distribution functions.
\end{enumerate}

\section{Application}\label{resultados}

The data analyzed in this study comprise daily maximum temperature and minimum relative humidity observations recorded in Brasília, Brazil, by the Brazilian National Institute of Meteorology (INMET), from January 2015 to December 2024.

\subsection{Descriptive Analysis of the Series}
\label{subsec:descritiva}
\begin{figure}[htbp]
\centering
\includegraphics[width=0.7\textwidth]{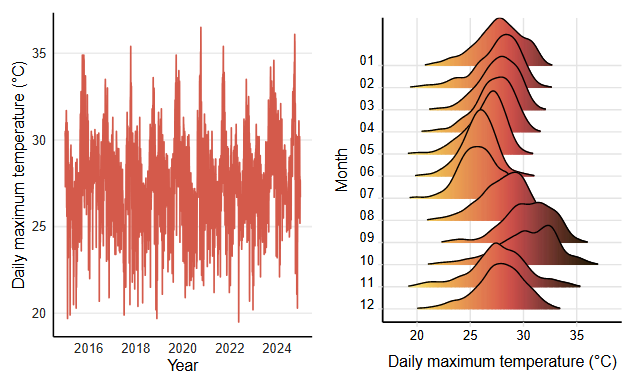}
\caption{Time series of daily maximum temperatures in $^\circ$C (left) and empirical density of maximum temperatures by month (right).}
\label{fig:descritivaT}
\end{figure}

\begin{figure}[htbp]
\centering
\includegraphics[width=0.7\textwidth]{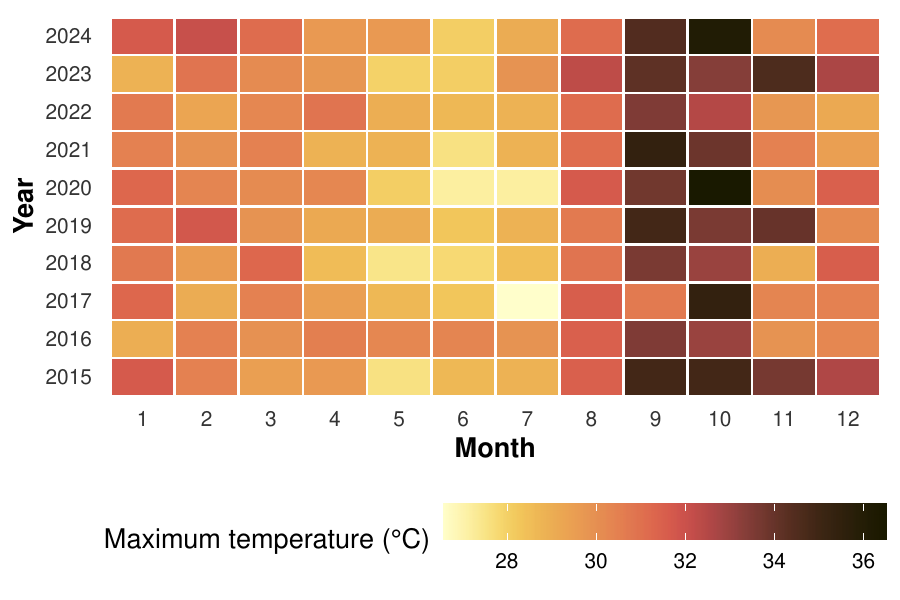}
\caption{Heatmap of daily maximum temperature ($^\circ$C).}
\label{fig:intensidadeT}
\end{figure}

\begin{figure}[htbp]
\centering
\includegraphics[width=0.40\textwidth]{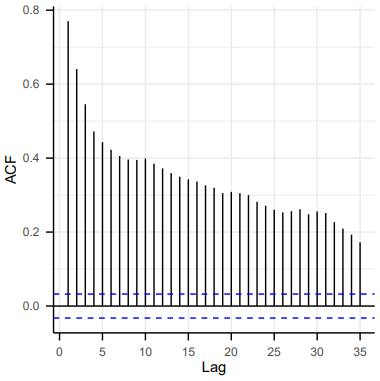}
\caption{Autocorrelation function of daily maximum temperature ($^\circ$C).}
\label{fig:acf}
\end{figure}

The descriptive analysis of daily maximum temperatures in Brasília over the studied period reveals seasonal components (Figures \ref{fig:descritivaT} and \ref{fig:intensidadeT}) and strong temporal dependence (Figure \ref{fig:acf}).

The left panel of Figure \ref{fig:descritivaT} shows that the maximum temperature series contains some extreme values exceeding 35 $^\circ$C.

The right panel of Figure \ref{fig:descritivaT} and Figure \ref{fig:intensidadeT} clearly reveal the seasonal structure and interannual variability of maximum temperatures in Brasília. The color gradient in Figure \ref{fig:intensidadeT} shows that the temperature regime of the region exhibits marked annual regularity, with recurring peaks during the final months of the dry season, particularly September and October, and more moderate values during the rainy season, which extends from December to March.

The seasonal pattern is evident from the recurring bluish blocks during the first half of each year, indicating lower mean maximum temperatures, generally ranging from 26 $^\circ$C to 29 $^\circ$C. By contrast, the central months of the dry season, particularly August, September, and October, consistently appear in warmer colors, reflecting maximum temperatures that frequently exceed 32 $^\circ$C and, in some years, approach 35 $^\circ$C.

The interannual analysis shows that, although the seasonal structure appears stable, there are notable variations among individual years. 

During the rainy months, from November to March, the densities are concentrated at lower values, often between 24 $^\circ$C and 28 $^\circ$C, with short tails indicating a lower occurrence of extreme heat. In contrast, the dry-season months, particularly from August to October, exhibit right-shifted distributions, with more days above 30 $^\circ$C and heavier tails, indicating episodes of intense heat. 
These months stand out for their right-shifted densities, consistent with the co-occurrence of the highest maximum temperatures and low relative humidity, a phenomenon typical of the end of the dry season in this city. 

In view of this phenomenon, the other climate variable considered here is relative humidity ($Y$).

\begin{figure}[htbp]
\centering
\includegraphics[width=0.7\textwidth]{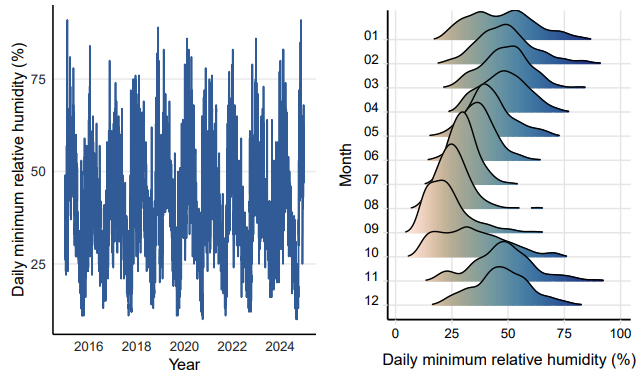}
\caption{Time series of daily minimum relative humidity in $\%$ (left) and empirical density of monthly minimum relative humidity (right).}
\label{fig:descritivaU}
\end{figure}

\begin{figure}[htbp]
\centering
\includegraphics[width=0.7\textwidth]{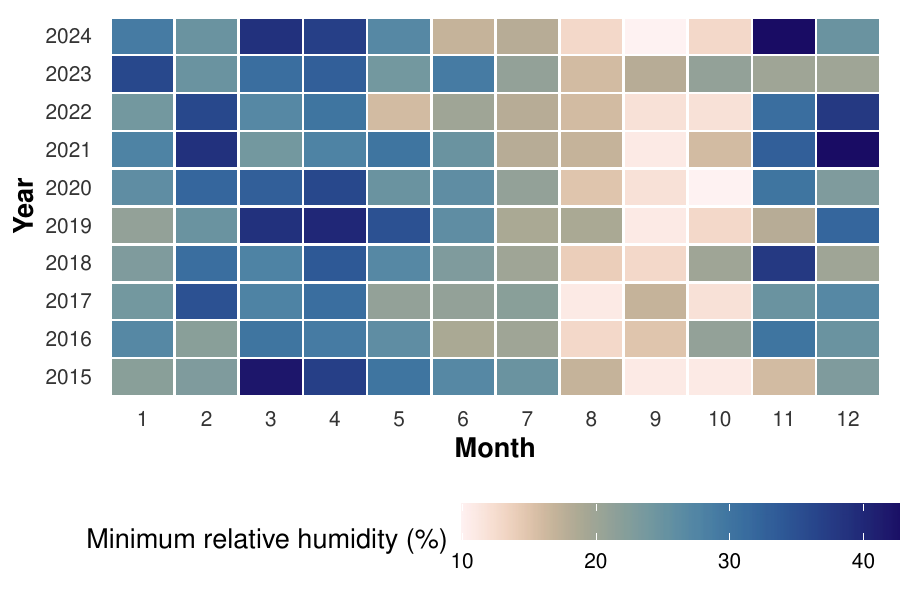}
\caption{Intensity map of daily minimum relative humidity (\%).}
\label{fig:mapaU}
\end{figure}

\begin{figure}[htbp]
\centering
\includegraphics[width=0.40\textwidth]{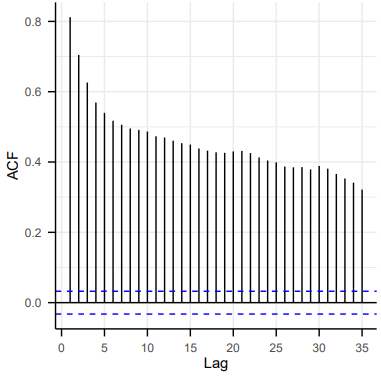}
\caption{Autocorrelation function of daily minimum relative humidity (\%).}
\label{fig:acfU}
\end{figure}

The time series (left panel of Figure \ref{fig:descritivaU}), covering ten years of daily observations, exhibits more pronounced cyclical behavior than that observed for maximum temperature. The alternation between extremely dry and humid periods follows a well-defined periodic pattern.

In the right panel of Figure \ref{fig:descritivaU}, the monthly densities are concentrated at high values, above 60\%, during the peak of the rainy season, from December to March. In contrast, August and September exhibit strongly left-shifted distributions, with modes between approximately 20\% and 30\% and tails extending below 15\%. July and October show intermediate distributions, characterizing transitional periods between the extremely dry and humid regimes.

The shades of blue in Figure \ref{fig:mapaU} confirm the pattern characterized by the alternation between the dry and rainy seasons, a feature of the tropical highland climate in Brasília. Thus, minimum relative humidity in Brasília ($Y$), like maximum temperature ($X$), exhibits strong seasonality, pronounced asymmetry, and high variability, all of which are strongly influenced by the season of the year.

\begin{figure}[htbp]
\centering
\includegraphics[width=0.9\textwidth]{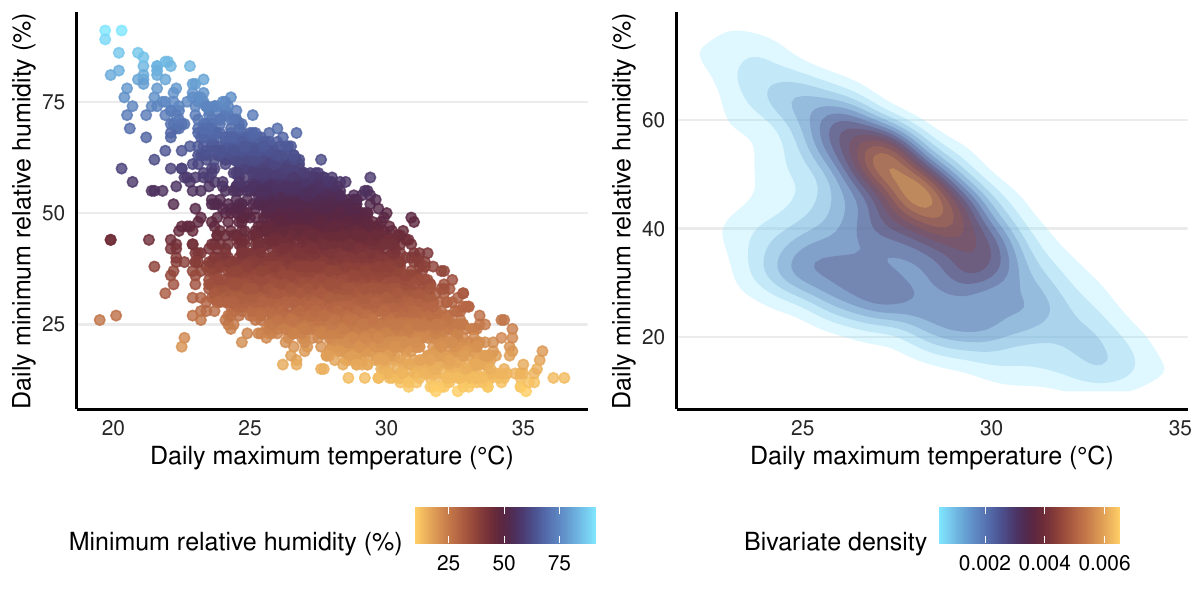}
\caption{Scatterplot of daily maximum temperature versus daily minimum relative humidity (left) and bivariate density plot (right).}
\label{dispersao}
\end{figure}

The joint assessment of the variables $X$ and $Y$ is initially conducted using the scatterplot presented in Figure \ref{dispersao}, which reveals a nonlinear relationship between the variables. At lower temperatures, minimum relative humidity exhibits high variability, taking both high and low values. In particular, within the approximate range of 20 $^\circ$C to 26 $^\circ$C, greater heterogeneity is observed in the distribution of minimum relative humidity. 
However, the density contours reveal substantial dispersion around the main core, indicating high joint variability between the variables. This pattern is consistent with the climate seasonality of the region, reflecting conditions ranging from humid days, characteristic of the rainy season, to extremely dry days, typical of the dry season.

\subsection{Return Level}
\label{subsec:nivel_ret}

In this subsection, the eight steps described in Subsection \ref{subsec:procedimento} are applied to determine the return levels associated with a series of extremes $\{(x_i,y_i)\}_{i=1}^{m}$ of $(X,Y)$. The marginal distributions of the extreme-value data are fitted using the BGEV distribution defined in \eqref{Fmarg}. As presented by \cite{otiniano2025revised}, the BGEV distribution was developed to model maxima and minima obtained from blocks of sufficiently large size $N$.

Thus, as described in Step 1 of Subsection \ref{subsec:procedimento}, the subsamples $\{x_i\}_{i=1}^{m}$ and $\{y_i\}_{i=1}^{m}$ were initially obtained, corresponding, respectively, to maximum temperatures and minimum relative humidity values, where $m=n/N$. The block size $N=64$ days was selected as the smallest size for which the Ljung--Box test did not reject the null hypothesis of no autocorrelation at the $5\%$ significance level. 
For this purpose, the Ljung--Box test was applied, and the smallest value of $N$ for which the test failed to reject the null hypothesis of no autocorrelation at the $5\%$ significance level was selected.

\begin{figure}[h!]
\includegraphics[width=0.5\textwidth]{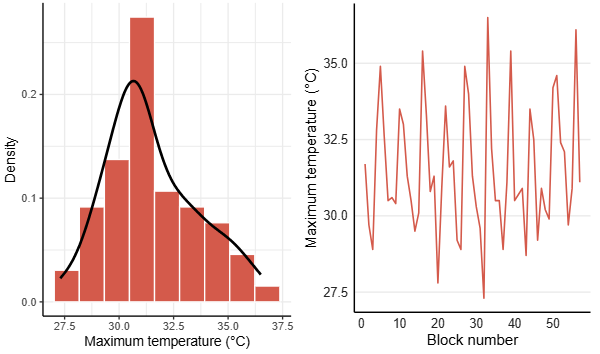} 
\includegraphics[width=0.5\textwidth]{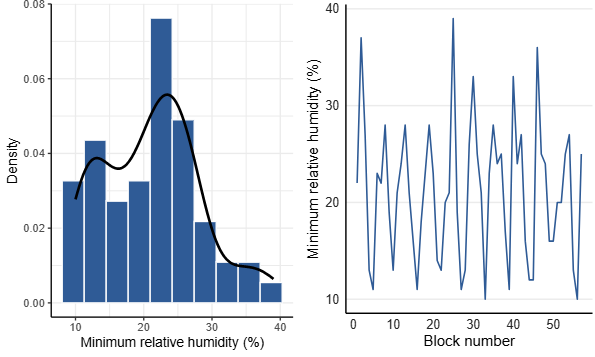}
\caption{Histogram and time series of $\{x_i\}_{i=1}^{m}$ in red, and histogram and time series of $\{y_i\}_{i=1}^{m}$ in blue.}
\label{fig:blocos}
\end{figure}

\begin{figure}[h!]
\centering
\includegraphics[width=0.7\textwidth]{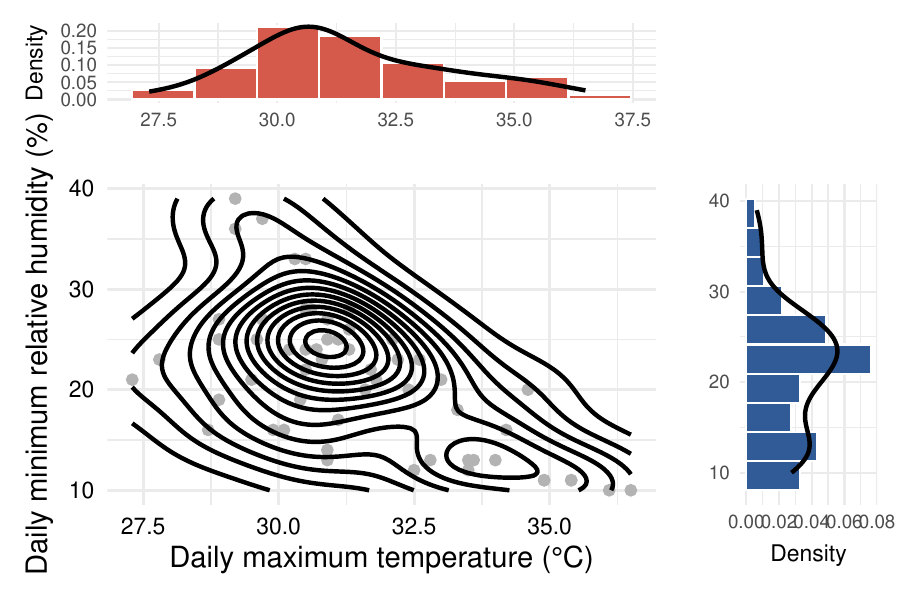}
\caption{Scatterplot of the maximum temperature and minimum relative humidity subsamples.}
\label{fig:scatter}
\end{figure}

In Figures \ref{fig:blocos} and \ref{fig:scatter}, the maximum temperature values show that the empirical density exhibits unimodal behavior, suggesting a degree of homogeneity in the temperature maxima over the period analyzed. Thus, a density compatible with the GEV distribution can be used to fit these data. 

The empirical density of minimum relative humidity suggests the coexistence of two distinct low-humidity regimes over the period analyzed. This behavior indicates greater heterogeneity in the lower extremes of the distribution, possibly associated with different atmospheric or seasonal conditions. In contrast to maximum temperature, whose distribution exhibits essentially unimodal behavior, the minimum relative humidity values exhibit a bimodal structure. Therefore, the BGEV distribution is more suitable than the GEV distribution for modelling these data.

In summary, in Step 2 of the procedure described in Subsection \ref{subsec:procedimento}, $\{x_i\}_{i=1}^{m}$ is fitted using the GEV distribution, whereas $\{y_i\}_{i=1}^{m}$ is fitted using the BGEV distribution. The \textit{fExtremes} package (version 4032.84) by \cite{pacotegev} was used to obtain the maximum likelihood estimates for the GEV distribution, while the \textit{bgev} package (version 0.1) by \cite{pacotebgev} was used to obtain the BGEV estimates. The parameter estimates for both the GEV and BGEV distributions are presented in Table \ref{estimmarginais}.

\begin{table}[h!]
\caption{Estimated parameters of the GEV and BGEV distributions.}
\centering
\small
\begin{tabular}{p{0.28\textwidth}ccccc}
\toprule
Variable & Distribution & $\xi$& $\mu$ & $\sigma$  & $\delta$\\
\midrule
Maximum temperature & GEV & -0.1235 & 30.6237 & 1.8401 & - \\
\midrule
Minimum relative humidity &GEV & -0.1616 & 18.2097 & 6.4909 & - \\
& BGEV & -0.0323 & 17.6569 & 14.5883 & 0.4020 \\
\bottomrule
\end{tabular}
\label{estimmarginais}
\end{table}

\begin{figure}[h!]
\centering
\includegraphics[width=0.85\textwidth]{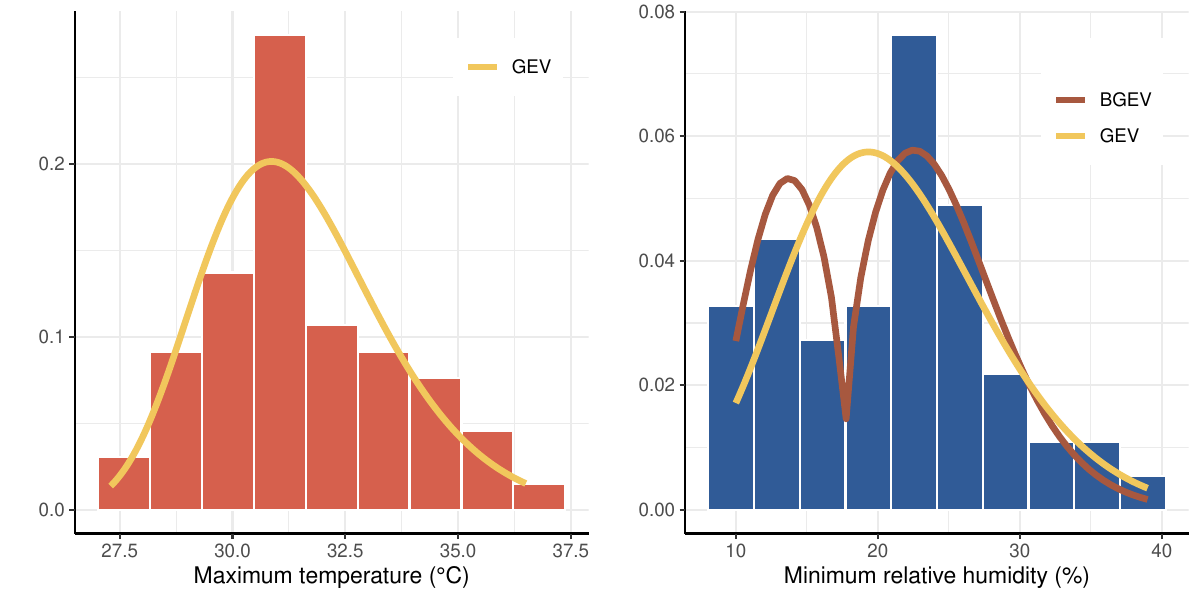}
\caption{Histograms of the GEV fit to maximum temperature and the GEV and BGEV fits to minimum relative humidity.}
\label{fig:fitmarg}
\end{figure}

The corresponding densities were generated using the parameters in Table \ref{estimmarginais} and are shown in Figure \ref{fig:fitmarg}. Clearly, minimum relative humidity is better fitted by the BGEV density than by the GEV density.

Having fitted the appropriate marginal distributions to each variable, we proceed to Steps 3 and 4 of the procedure described in Subsection \ref{subsec:procedimento}.

Once the data were transformed to $u_i=\{ F_X(x_i)\}_{i=1}^{m}, F_X= GEV$ and $\{v_i=F_Y(y_i)\}_{i=1}^{m}, F_Y=BGEV$, a copula $C$ is fitted to $\{(u_i, v_i)\}_{i=1}^{m}$.

The \textit{VineCopula} package (version 2.6.1) by \cite{pacotecopula}, implemented in R, was used to test 28 copula families. 
The most appropriate model was selected based on the lowest AIC and BIC values. As shown in Table \ref{AICBICCOP}, the best-fitting copula was the Tawn Type I copula rotated by 270°.

\begin{table}[h!]
\caption{Log-likelihood, AIC, and BIC values for the candidate copula models fitted to the block maxima of maximum temperature and block minima of minimum relative humidity (\%) in Brasília.}
\centering
\small
\begin{tabular}{rlrrr}
  \toprule
 & Copula & loglik & AIC & BIC \\ 
  \midrule
1 & Independence & 0.00 & 2.00 & 4.04 \\ 
  2 & Gaussian & 14.75 & -27.51 & -25.46 \\ 
  3 & T-Student & 14.65 & -27.31 & -25.27 \\ 
  4 & Clayton & -0.00 & 2.01 & 4.05 \\ 
  5 & Clayton rotated 90° & 22.60 & -43.19 & -41.15 \\ 
  6 & Clayton rotated 180° & -0.00 & 2.01 & 4.05 \\ 
  7 & Clayton rotated 270° & 5.27 & -8.54 & -6.49 \\ 
  8 & Gumbel & -0.01 & 2.01 & 4.06 \\ 
  9 & Gumbel rotated 90° & 8.44 & -14.89 & -12.84 \\ 
  10 & Gumbel rotated 180° & -0.01 & 2.01 & 4.06 \\ 
  11 & Gumbel rotated 270° & 19.39 & -36.78 & -34.74 \\ 
  12 & Frank & 14.33 & -26.66 & -24.61 \\ 
  13 & Joe & -0.00 & 2.01 & 4.05 \\ 
  14 & Joe rotated 90° & 3.32 & -4.63 & -2.59 \\ 
  15 & Joe rotated 180° & -0.00 & 2.01 & 4.05 \\ 
  16 & Joe rotated 270° & 22.65 & -43.29 & -41.25 \\ 
  17 & BB1 rotated 90° & 22.59 & -43.18 & -41.14 \\ 
  18 & BB1 rotated 270° & 19.38 & -36.76 & -34.72 \\ 
  19 & BB6 rotated 90° & 8.43 & -14.87 & -12.83 \\ 
  20 & BB6 rotated 270° & 22.64 & -43.29 & -41.24 \\ 
  21 & BB7 rotated 90° & 22.59 & -43.18 & -41.14 \\ 
  22 & BB7 rotated 270° & 22.64 & -43.29 & -41.24 \\ 
  23 & BB8 rotated 90° & 11.93 & -21.86 & -19.82 \\ 
  24 & BB8 rotated 270° & 22.65 & -43.29 & -41.25 \\ 
  25 & Tawn Type I rotated 90° & 5.55 & -9.09 & -7.05 \\ 
  \textbf{26} & \textbf{Tawn Type I rotated 270°} & \textbf{24.49} & \textbf{-46.98} & \textbf{-44.94} \\ 
  27 & Tawn Type II rotated 90° & 10.59 & -19.17 & -17.13 \\ 
  28 & Tawn Type II rotated 270° & 14.55 & -27.09 & -25.05 \\ 
\bottomrule
\end{tabular}
\label{AICBICCOP}
\end{table}

\begin{figure}[htbp]
\centering
\includegraphics[width=0.9\textwidth]{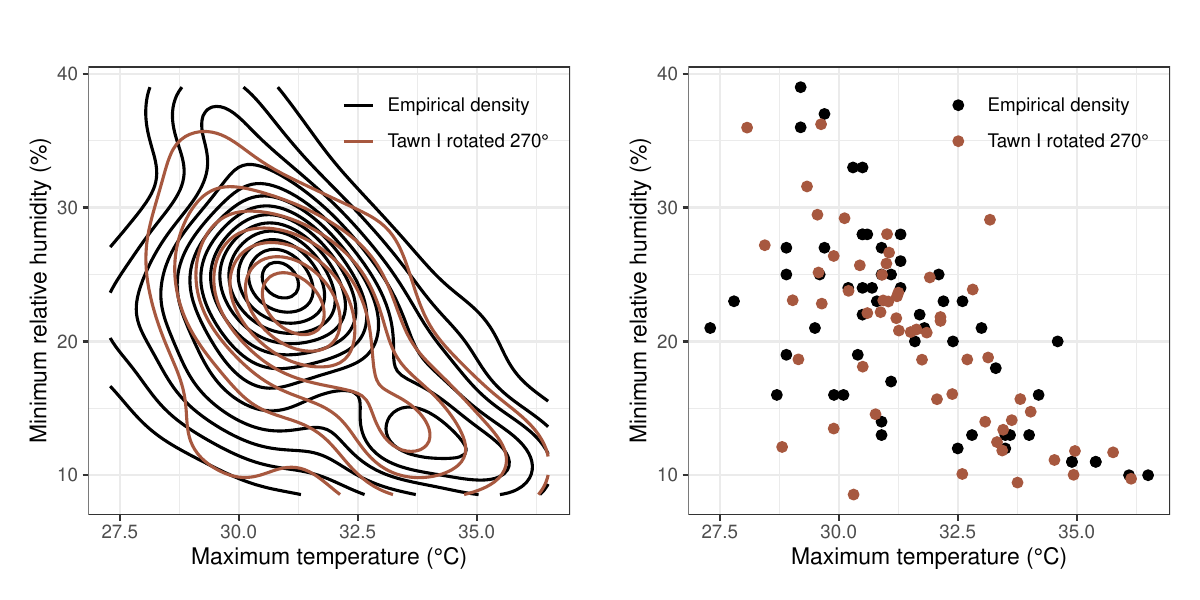}
\caption{Fitted model in red and data $\{(x_i, y_i)\}_{i=1}^{m}$ in black. Smoothed curves and contour lines of the fitted model (left), and empirical and simulated data (right).}
\label{copula}
\end{figure}

The left panel of Figure \ref{copula} shows that the model satisfactorily reproduced the joint structure of the data, clearly capturing the two characteristic peaks arising from the bimodality of minimum relative humidity. The contour lines exhibited an alignment and shape consistent with the empirical density, demonstrating that the selected copula was able not only to represent the overall dependence between the variables but also to adequately capture the two distinct regions of concentration in the bivariate distribution.

The scatterplot (right) reinforces this conclusion, as the simulated point cloud exhibits a pattern similar to that observed empirically, preserving the overall shape, the relative concentration of the blocks, and the consistency of the extremes. This performance indicates that the fitted copula accurately reproduces the geometry of the relationship between maximum temperature and minimum relative humidity, visually validating the fit and confirming that the model is appropriate for describing the dependence between these extreme meteorological variables.

In addition to the 2-dimensional analyses, 3-dimensional plots were produced for the empirical density and the density modeled using the selected copula. A plot of the modeled density with both margins following GEV distributions was also produced for comparison. These 3D plots provide an even clearer and more detailed visualization of the joint structure of the extremes, allowing a direct assessment of the model's ability to reproduce the complexity observed in the data.

\begin{figure}[htbp]
\centering
\includegraphics[width=1\textwidth]{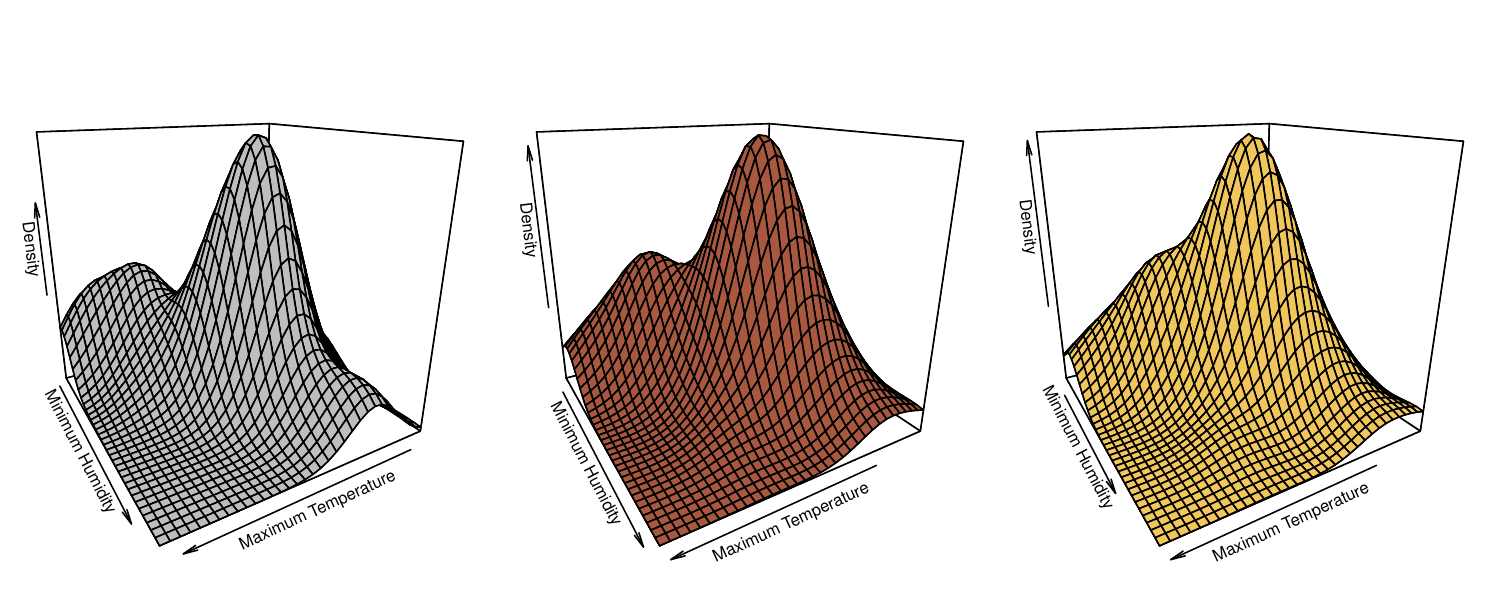}
\caption{Empirical density (left), fitted density with GEV- BGEV margins (center), and fitted density with GEV - GEV margins (right).}
\label{3d}
\end{figure}

A comparison of the three surfaces shows that the two characteristic peaks of the empirical bivariate distribution were captured with excellent accuracy by the model with GEV - BGEV margins, which was not observed for the model with GEV - GEV margins (see Figure \ref{3d}). The empirical surface exhibits two well-defined peaks corresponding to distinct regimes of extreme behavior, and the surface modeled with GEV - BGEV margins reproduces these peaks with remarkable accuracy in terms of both location and relative intensity. This agreement demonstrates that the selected copula, combined with well-fitted margins, was able to represent not only the overall dependence between the variables but also the geometry of the joint distribution, a fundamental aspect of extreme-value studies. The model with GEV - GEV margins clearly provided an inferior fit, as it virtually ignored the smaller second peak observed in the empirical density. This lack of fit results in a severe overestimation error, particularly in extreme-value analysis, because it fails to account precisely for the subset of data containing the more extreme values in region $I$, which is the focus of this study. 

Thus, the model's ability to accurately capture both peaks reinforces the robustness of the adopted approach and confirms that the dependence structure was adequately estimated, ensuring a more realistic and reliable representation of joint temperature and humidity extremes.

The impact of inadequately fitting the marginal distributions on the estimation of risk measures can be clearly observed in the \textbf{return-level values} presented in Table \ref{tab_var_final}. In particular, the comparison between the models with GEV - BGEV and GEV - GEV margins reveals substantial differences in the joint characterization of extreme events involving high maximum temperatures and low minimum relative humidity.

\begin{table}[h!]
\caption{Estimated joint return-level pairs for maximum temperature
($^\circ$C) and minimum relative humidity (\%) in Brasília under the
GEV--BGEV and GEV--GEV marginal specifications for selected return
periods.}
\label{tab_var_final}

\centering
\small
\renewcommand{\arraystretch}{1.15}
\setlength{\tabcolsep}{6pt}

\begin{tabular}{@{}ccccc@{}}
\toprule

\multirow{3}{*}{\shortstack{Return period\\(in days)}}
& \multicolumn{4}{c}{Return-Level Estimation} \\

\cmidrule(lr){2-5}

& \multicolumn{2}{c}{\textbf{(GEV, BGEV)}}
& \multicolumn{2}{c}{\textbf{(GEV, GEV)}} \\

\cmidrule(lr){2-3}
\cmidrule(lr){4-5}

& Max. temp.
& Min. rel. humidity
& Max. temp.
& Min. rel. humidity \\

\midrule
\addlinespace[-0.4em]
\midrule

65   & 28.40 & 43.52 & 28.40 & 46.19 \\
100  & 30.68 & 34.98 & 30.68 & 37.54 \\
365  & 33.41 & 23.38 & 33.41 & 23.48 \\
500  & 33.91 & 21.93 & 33.91 & 21.93 \\
730  & 34.46 & 17.54 & 34.46 & 18.49 \\
1825 & 35.66 & 12.61 & 35.66 & 13.91 \\
3650 & 36.49 & 11.38 & 36.49 & 12.22 \\

\bottomrule
\end{tabular}

\end{table}

Table \ref{tab_var_final} presents the return levels associated with the joint probability of extreme events, defined as the maximum temperature exceeding a given threshold while minimum relative humidity simultaneously falls below another threshold. The return period, expressed in days, should be interpreted as the average time interval between events of this type. For example, for a return period of 65 days, the maximum temperature is expected, on average, to exceed 28.40 $^\circ$C while minimum relative humidity falls below 43.52\%, based on the GEV–BGEV fit.

The results obtained with the GEV–BGEV model, which provided a better fit to the marginal distributions, reveal coherent and physically plausible behavior of the joint return levels. As the return period increases, the values associated with maximum temperature gradually increase, whereas those associated with minimum relative humidity become progressively lower, reflecting the increasingly extreme nature of the events considered. This pattern is consistent with extreme-value theory and with the climate dynamics expected for rare events.

The discrepancy between the GEV–BGEV and GEV–GEV fits becomes evident when directly comparing the return-level values associated with minimum relative humidity. For short return periods, such as 65 days, the GEV–GEV model estimates a minimum humidity threshold of 46.19\%, whereas the GEV–BGEV fit yields 43.52\%, corresponding to a relative difference of approximately 6\%. This overestimation persists even for longer return periods: for 365 days, for example, the estimated values are 23.48\% (GEV–GEV) versus 23.38\% (GEV–BGEV), and for 3650 days, 12.22\% versus 11.38\%, representing relative differences of approximately 7.4\%. Although small in percentage terms in some cases, these discrepancies are systematic and occur precisely in the lower tail of the distribution, where small absolute variations correspond to substantial changes in the severity of extreme events, reinforcing the adverse impact of a poor marginal fit on the estimation of joint return levels.

This overestimation of the return level for minimum relative humidity is particularly problematic in scientific studies of climate extremes. In practical applications, such as environmental risk assessment, urban planning, public health, and extreme-event management, underestimating the severity of critical episodes may lead to overly optimistic conclusions and, consequently, inadequate mitigation strategies. 

Therefore, the comparison between the models shows that the indiscriminate use of the GEV distribution for all margins, without a careful assessment of its fit, may introduce significant biases into the estimation of extreme-risk measures. The results reinforce the importance of appropriate marginal modelling, such as that provided by the BGEV model for minimum relative humidity, particularly when the objective is the reliable inference of joint return levels in extreme-value studies.

\section{Final Considerations} \label{final}
This study estimated joint return levels associated with maximum temperature and minimum relative humidity in Brasília, Brazil, by integrating the BGEV distribution, proposed as a new approach to extreme-value modelling, with copula-based dependence-structure modelling.

The BGEV distribution provided a better representation of minimum relative humidity extremes. By accommodating multiple peaks in the marginal distribution, the BGEV provided more consistent and realistic fits, reducing bias in the estimation of extreme quantiles and risk measures. The results showed that the use of poorly fitted models may lead to the systematic overestimation of return levels associated with minimum relative humidity, which, in practical applications, may result in misleading interpretations of the severity and recurrence of extreme events.

Joint modelling using copulas made it possible to flexibly capture the dependence between maximum temperature and minimum relative humidity, given the marginal distributions adopted. The evaluation of different copula families revealed that the choice of dependence structure directly influences joint risk estimation, particularly in the tails of the distribution, where simultaneous extreme events are most relevant.

From a practical perspective, the results of this study have direct implications for environmental planning, climate risk management, and public policy development in the Federal District. The accurate identification of thresholds associated with the simultaneous occurrence of high temperatures and low humidity levels is essential for supporting strategies to mitigate impacts on public health, water security, and the conservation of the Cerrado biome. Furthermore, the proposed methodology provides a robust statistical framework that can be applied to other regions and climate variables, contributing to a more accurate assessment of the risks associated with climate change.

Consequently, this study reinforces the importance of carefully selecting statistical models for the analysis of climate extremes, demonstrating that more flexible approaches, such as the BGEV distribution combined with copula modelling, are essential for capturing the complexity of real-world data. By highlighting the effects of inadequate model fits on the estimation of risk measures, this study makes a significant contribution to the literature on extreme value analysis applied to climate data and provides relevant methodological insights for future research under scenarios of increasingly intense climate extremes.

As a direction for future research, the proposed model could be extended by incorporating covariates, thereby accounting for meteorological and environmental factors that may influence the joint behavior of maximum temperature and minimum relative humidity and yielding more accurate and informative return-level estimates.

\vspace{1.2cm}

\noindent\rule{0.4\textwidth}{0.35pt}

\noindent \emph{\textbf{Data availability.}}
The data are publicly available on the official website of the Brazilian
National Institute of Meteorology (INMET),
\url{https://bdmep.inmet.gov.br/}.

\vspace{0.45cm}

\noindent \emph{\textbf{Author contributions.}}
BGA: conceptualization, writing (original draft preparation), data curation;
CSB: methodology, data curation, validation;
CEGO: supervision, methodology, writing (review and editing);
FM: writing (review and editing);
EPB: writing (review and editing).

\vspace{0.45cm}

\noindent \emph{\textbf{Acknowledgments.}}
This research received scholarship support from the Coordination for the
Brazilian Improvement of Higher Education Personnel (CAPES), the Brazilian
National Council for Scientific and Technological Development (CNPq) and the
University of Brasília (UnB) through its Research Support Program.

\vspace{1cm}

\newpage
\bibliographystyle{plainnat}
\bibliography{bibliography}

@book{joe2014dependence,
  title={Dependence modeling with copulas},
  author={Joe, Harry},
  year={2014},
  publisher={CRC press}
}

@techreport{joe1996estimation,
  title={The estimation method of inference functions for margins for multivariate models},
  author={Joe, Harry and Xu, James Jianmeng},
  institution = {Department of Statistics, University of British Columbia},
  type        = {Technical Report},
  number      = {166},
    address     = {Vancouver, Canada},
  year={1996}
}

@inproceedings{sklar1959fonctions,
  title={Fonctions de r{\'e}partition {\`a} n dimensions et leurs marges},
  author={Sklar, M},
  booktitle={Annales de l'ISUP},
  volume={8},
  pages={229--231},
  year={1959}
}

@article{otiniano2025revised,
  title={A Revised Bimodal Generalized Extreme Value Distribution: Theory and Climate Data Application},
  author={Otiniano, Cira EG and Lisboa, Mathews NS and Ribeiro, Terezinha KA},
  journal={Entropy},
  volume={27},
  number={7},
  pages={749},
  year={2025},
  publisher={MDPI}
}

@book{salvadori2007extremes,
  title={Extremes in nature: an approach using copulas},
  author={Salvadori, Gianfausto and De Michele, Carlo and Kottegoda, Nathabandu T and Rosso, Renzo},
  volume={56},
  year={2007},
  publisher={Springer Science \& Business Media}
}

@book{IPCC2013,
  author    = {{IPCC}},
  title     = {Climate Change 2013: The Physical Science Basis},
  subtitle  = {Contribution of Working Group I to the Fifth Assessment Report of the Intergovernmental Panel on Climate Change},
  publisher = {Cambridge University Press},
  address   = {Cambridge, United Kingdom and New York, NY, USA},
  year      = {2013},
  note      = {Edited by Stocker, T. F., Qin, D., Plattner, G.-K., Tignor, M., Allen, S. K., Boschung, J., Nauels, A., Xia, Y., Bex, V., and Midgley, P. M.; 1535 pp.}
}

@misc{WMO2026,
  author      = {World Meteorological Organization},
  title       = {State of the Global Climate 2025},
  year        = {2026},
  institution = {World Meteorological Organization (WMO)},
  address     = {Geneva, Switzerland},
  month       = {March},
  note        = {Published on 23 March 2026}
}

@article{ETIOPIA,
author = {Degfie Teku},
title = {Navigating climate uncertainty: a comprehensive review of climatic variabilities and extreme events on environmental, socio-economic, and livelihood dimensions in Ethiopia with adaptation strategies},
journal = {All Earth},
volume = {37},
number = {1},
pages = {1--30},
year = {2025},
publisher = {Taylor \& Francis}
}

@article{INDIA,
title = {Temperature projections and heatwave attribution scenarios over India: A systematic review},
journal = {Heliyon},
volume = {10},
number = {4},
pages = {e26431},
year = {2024},
issn = {2405-8440},
doi = {https://doi.org/10.1016/j.heliyon.2024.e26431},
url = {https://www.sciencedirect.com/science/article/pii/S2405844024024629},
author = {Khaiwal Ravindra and Sanjeev Bhardwaj and Chhotu Ram and Akshi Goyal and Vikas Singh and Chandra Venkataraman and Subhash C. Bhan and Ranjeet S. Sokhi and Suman Mor}
}

@article{ANTARTICA,
author = {Wanderley, Henderson and Justino, Flávio and Sediyama, Gilberto},
year = {2016},
month = {06},
pages = {114-121},
title = {Tendência da Temperatura e Precipitação na Península Antártica},
volume = {31},
journal = {Revista Brasileira de Meteorologia},
doi = {10.1590/0102-778631220140146}
}

@incollection{AFRICA,
  author       = {{International Monetary Fund}},
  title        = {Adaptação às Alterações Climáticas na África Subsariana},
  booktitle    = {Perspectivas Económicas Regionais: África Subsariana},
  year         = {2020},
  publisher    = {International Monetary Fund},
  address      = {Washington, DC},
  month        = {April},
  note         = {Capítulo 2}
}

@article{GEV_BRASIL,
author = {Souza, Amaury and Medeiros, Elias and de Oliveira-Júnior, José and Kumar, Vikram and Gautam, Sneha and Bezerra, Aline},
year = {2025},
month = {01},
pages = {},
title = {Analyzing Maximum Temperature Trends and Extremes in Brazil: A Study of Climate Variability and Anthropogenic Influences from 1960 to 2020},
journal = {Aerosol Science and Engineering},
doi = {10.1007/s41810-025-00288-2}
}

@article{Clima_Brasil,
author = {Libonati, Renata and Geirinhas, João L. and Silva, Patrícia S. and Monteiro dos Santos, Djacinto and Rodrigues, Julia A. and Russo, Ana and Peres, Leonardo F. and Narcizo, Luiza and Gomes, Monique E. R. and Rodrigues, Andreza P. and DaCamara, Carlos C. and Pereira, José Miguel C. and Trigo, Ricardo M.},
title = {Drought–heatwave nexus in Brazil and related impacts on health and fires: A comprehensive review},
journal = {Annals of the New York Academy of Sciences},
volume = {1517},
number = {1},
pages = {44-62},
year = {2022},
doi = {https://doi.org/10.1111/nyas.14887},
url = {https://nyaspubs.onlinelibrary.wiley.com/doi/abs/10.1111/nyas.14887}
}

@techreport{PBMC2014,
  author       = {{PBMC}},
  title        = {Base Científica das Mudanças Climáticas: Contribuição do Grupo de Trabalho 1 ao Primeiro Relatório de Avaliação Nacional do Painel Brasileiro de Mudanças Climáticas},
  year         = {2014},
  institution  = {PBMC},
  address      = {Rio de Janeiro, RJ},
  volume       = {1}
}

@techreport{Menezes2016,
  author       = {Leila Soraya Menezes and Sin Chan Chou and Josefa Morgana Viturino de Almeida and Saulo Aires Souza and Wagner de Aragão Bezerra and Lineu Neiva Rodrigues and Carlos Henrique Eça D'Almeida Rocha},
  title        = {Mudanças Climáticas no DF e RIDE: Detecção e Projeções das Mudanças Climáticas para o Distrito Federal e Região Integrada de Desenvolvimento do DF e Entorno},
  institution  = {Secretaria do Meio Ambiente do Distrito Federal},
  year         = {2016},
  address      = {Brasília, DF},
  note         = {Nota técnica endereçada aos formuladores de políticas públicas e tomadores de decisão},
  isbn         = {978-85-68931-03-5}
}

@article{Bayer2015,
  author  = {Bayer, Débora Missio and Bayer, Fábio Mariano},
  title   = {Previsão da Umidade Relativa do Ar de Brasília por Meio do Modelo Beta Autorregressivo de Médias Móveis},
  journal = {Revista Brasileira de Meteorologia},
  volume  = {30},
  number  = {3},
  pages   = {319--326},
  year    = {2015},
  doi     = {10.1590/0102-778620130645}
}

@article{Cordeiro2020,
  author  = {Cordeiro, Shayane dos Santos and Otiniano, Cira E. Guevara},
  title   = {Modelos Bivariados de Eventos Climáticos Extremos do DF},
  journal = {Revista SODEBRAS},
  volume  = {15},
  number  = {180},
  pages   = {58--63},
  year    = {2020}
}

@misc{Correio2025,
  author       = {{Correio Braziliense}},
  title        = {Umidade pode baixar para 10\% e INMET emite alerta vermelho no DF},
  year         = {2025},
  url          = {https://www.correiobraziliense.com.br/cidades-df/2025/09/7259298-umidade-pode-baixar-para-10-e-inmet-emite-alerta-vermelho-no-df.html},
  note         = {Acesso em: 07 maio 2026}
}

@article{vandenberghe2012joint,
  title={Joint return periods in hydrology: a critical and practical review focusing on synthetic design hydrograph estimation.},
  author={Vandenberghe, S and Van den Berg, MJ and Gr{\"a}ler, B and Petroselli, A and Grimaldi, S and De Baets, B and Verhoest, NEC},
  journal={Hydrology \& Earth System Sciences Discussions},
  volume={9},
  number={5},
  year={2012}
}

@article{graler2013multivariate,
  title={Multivariate return periods in hydrology: a critical and practical review focusing on synthetic design hydrograph estimation},
  author={Gr{\"a}ler, Benedikt and van den BERG, M J and Vandenberghe, Sander and Petroselli, Andrea and Grimaldi, Salvatore and De Baets, Bernard and Verhoest, NEC},
  journal={Hydrology and Earth System Sciences},
  volume={17},
  number={4},
  pages={1281--1296},
  year={2013},
  publisher={Copernicus Publications G{\"o}ttingen, Germany}
}

@article{marengo2025clima,
  title={Clima: extremos e desastres},
  author={MARENGO, J},
  journal={Revista ClimaCom},
  volume={12},
  number={28},
  pages={1--21},
  year={2025}
}

@article{shiau2003return,
  title={Return period of bivariate distributed extreme hydrological events},
  author={Shiau, JT},
  journal={Stochastic environmental research and risk assessment},
  volume={17},
  number={1},
  pages={42--57},
  year={2003},
  publisher={Springer}
}

@Manual{pacotegev,
    title = {fExtremes: Rmetrics - Modelling Extreme Events in Finance},
    author = {Diethelm Wuertz and Tobias Setz and Yohan Chalabi},
    year = {2023},
    note = {R package version 4032.84},
    url = {https://CRAN.R-project.org/package=fExtremes},
    doi = {10.32614/CRAN.package.fExtremes},
  }

@Manual{pacotebgev,
    title = {bgev: Bimodal GEV Distribution with Location Parameter},
    author = {Cira Otiniano and Yasmin Lirio},
    year = {2024},
    note = {R package version 0.1},
    url = {https://CRAN.R-project.org/package=bgev},
    doi = {10.32614/CRAN.package.bgev},
  }

@Manual{pacotecopula,
    title = {VineCopula: Statistical Inference of Vine Copulas},
    author = {Thomas Nagler and Ulf Schepsmeier and Jakob Stoeber and Eike Christian Brechmann and Benedikt Graeler and Tobias Erhardt},
    year = {2025},
    note = {R package version 2.6.1},
    url = {https://CRAN.R-project.org/package=VineCopula},
    doi = {10.32614/CRAN.package.VineCopula},
  }

\end{document}